# Primrose: Selecting Container Data Types by Their Properties

Xueying Qin[a], Liam O'Connor[a], and Michel Steuwer[a]

a    The University of Edinburgh, Scotland, United Kingdom

**Abstract**
**Context**    Container data types are ubiquitous in computer programming, enabling developers to efficiently store and process collections of data with an easy-to-use programming interface. Many programming languages offer a variety of container implementations in their standard libraries based on data structures offering different capabilities and performance characteristics.
**Inquiry**    Choosing the *best* container for an application is not always straightforward, as performance characteristics can change drastically in different scenarios, and as real-world performance is not always correlated to theoretical complexity.
**Approach**    We present Primrose, a language-agnostic tool for selecting the best performing valid container implementation from a set of container data types that satisfy *properties* given by application developers. Primrose automatically selects the set of valid container implementations for which the *library specifications*, written by the developers of container libraries, satisfies the specified properties. Finally, Primrose ranks the valid library implementations based on their runtime performance.
**Knowledge**    With Primrose, application developers can specify the expected behaviour of a container as a type refinement with *semantic properties*, e.g., if the container should only contain unique values (such as a `set`) or should satisfy the LIFO property of a `stack`. Semantic properties nicely complement *syntactic properties* (i.e., traits, interfaces, or type classes), together allowing developers to specify a container's programming interface *and* behaviour without committing to a concrete implementation.
**Grounding**    We present our prototype implementation of Primrose that preprocesses annotated Rust code, selects valid container implementations and ranks them on their performance. The design of Primrose is, however, language-agnostic, and is easy to integrate into other programming languages that support container data types and traits, interfaces, or type classes. Our implementation encodes properties and library specifications into verification conditions in Rosette, an interface for SMT solvers, which determines the set of valid container implementations. We evaluate Primrose by specifying several container implementations, and measuring the time taken to select valid implementations for various combinations of properties with the solver. We automatically validate that container implementations conform to their library specifications via property-based testing.
**Importance**    This work provides a novel approach to bring abstract modelling and specification of container types directly into the programmer's workflow. Instead of selecting concrete container implementations, application programmers can now work on the level of specification, merely stating the behaviours they require from their container types, and the best implementation can be selected automatically.

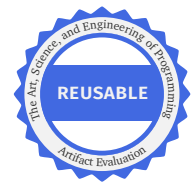

**Primrose: Selecting Container Data Types by Their Properties**

## 1 Introduction

Container data types, such as sets, lists, and trees, represent collections of data ubiquitous in everyday programming [11]. Virtually all programming languages provide a variety of different container implementations in their standard libraries.

Much work has been done to design better abstractions, improve performance and verify correctness for container data types. However, a crucial problem for application developers using containers still exists: when choosing a container data type, application developers are forced to select a concrete implementation that comes with certain theoretical complexity and practical performance tradeoffs.

For example, consider representing a mathematical set, i.e., where each element should occur at most once. In C++, we must choose between std::set, usually implemented as red-black trees [4], and std::unordered_set, implemented as a hash table. The hash-based implementation was added to the C++ standard in 2011, as the C++ standard has strict complexity requirements preventing the ordinary std::set to be implemented as the (often faster) hash table. Many blog posts and discussions [1, 3, 7, 25, 38] report on the performance of various C++ containers, showing the community's interest and the need for external guidance that the language itself does not provide.

In other languages, the situation is similar. Rust provides two container implementations, HashSet and BTreeSet, expecting application developers to make an explicit choice between them. Scala's complex collection library features abstract interfaces, such as the Set trait, abstracting over many implementations such as HashSet and TreeSet. But when creating an instance of Set, a default HashSet implementation is chosen regardless of the suitability of this implementation choice for the usage pattern of the application.

These examples demonstrate a general problem: Application developers are forced to *overspecify*, by having to select a concrete implementation, where we generally would like application developers to be shielded from low-level implementation details. Application developers should primarily care about the *abstract behaviour* of the containers in their application, and not how this is achieved. The compiler, or a dedicated tool, should identify those containers that satisfy their functional requirements, and select the best implementation automatically.

In this paper, we propose such an automated tool: Primrose, which allows application developers to specify the expected behaviours and programming interfaces of containers as *properties*. *Syntactic properties* specify the required programming interface of the container and are expressed as traits of the underlying programming language. *Semantic properties* specify the expected behaviour of the container and are written as logical predicates used as refinements of the container type. Primrose automatically selects the set of valid implementations for which the *library specifications*, written by the library developers as pre- and post-conditions of the container operations, satisfy the specified syntactic and semantic properties using an SMT solver. Finally, Primrose ranks the valid library implementations based on their runtime performance.

To select the best container implementation, first those container implementations which meet the functional requirements of the application developer must be determined, and then those valid container implementations must be evaluated based on non-functional requirements. While Primrose does include functionality for ranking based on benchmarks, the focus of this paper is on the first of these two problems. There are many existing sophisticated techniques for selecting based on non-functional requirements, and they are highly complementary with Primrose.

In this work, we apply verification and formal methods techniques, including refinement types, formal library specifications, and SMT solvers, in an innovative way to raise the level of abstraction for developers, freeing them from the burden of choosing container implementations, and opening up the possibility to automatically improve the performance of applications.

To summarize, this paper makes the following contributions:

- We present Primrose (Section 3), a language-agnostic tool for selecting valid container implementations (Section 6) based on *properties* (Section 4) describing their behaviours and programming interfaces and ranking them based on performance.
- We show a new application of refinement types (Section 4) not—as previous work did—for verification purposes, but to raise the level of abstraction for developers and to improve the runtime performance of applications with container data types.
- We develop a new methodology to specify container libraries (Section 5), amenable to our selection process, making use of existing formal methods work such as data abstraction and Hoare logic.
- We show the feasibility of Primrose, selecting container implementations that satisfy various properties from a Rust library of eight container types with library specifications. We validate container implementations against specifications and evaluate the efficiency of the selection process (Section 7).

## 2 Motivation

Suppose as part of a larger application we want to find and store all the elements of a larger collection, but without duplicates. We might, for example, use the result of this function to count the number of unique elements or process the elements further, now with the guarantee that each element in the returned collection is unique.

An easy way to implement this is to return a container that only permits unique elements. We might think of a *set*, but as discussed in Section 1, this requires a choice: Which implementation of the abstract idea of a mathematical set should we use?

Figure 1a shows a Rust code snippet computing a container uniqueElements that contains the unique elements of the original input sequence. The application developer must choose a concrete container implementation, such as HashSet in line 1, but other valid choices would be Rust's BTreeSet (line 2), or perhaps a custom UniqueVect (line 3) container, which stores all elements in a vector but ensures there are no duplicates, or some other FancySetImplementations (line 4). Whether a container implementation is *valid* is determined by the application developer's *functional requirements*. Our uniqueness requirement, for example, is not met by the Rust HashMultiSet (line 5).

Many programming techniques exist to abstract over multiple concrete implementations of a general concept. In object-oriented languages, *abstract classes* enable hiding multiple implementations behind a common interface. Similar features exist in other

**Primrose: Selecting Container Data Types by Their Properties**

```rust
1  type Set<I> = HashSet<I>;                                Rust
2  // type Set<I> = BTreeSet<I>;
3  // type Set<I> = UniqueVect<I>;
4  // type Set<I> = FancySetImpl<I>;
5  // type Set<I> = HashMultiSet<I>; ???
6
7  let mut uniqueElements = Set::new();
8  for val in input.iter() {
9      uniqueElements.insert(val); }
```

```
1  property unique {                                     Primrose
2    \c -> (for-all-elems (\a ->
3                 (unique-count? a c)) c) };
4  type UniqueCon<I> = {
5    c <: ContainerT | unique c };
                                                            Rust
7  let mut uniqueElements = UniqueCon::new();
8  for val in input.iter() {
9      uniqueElements.insert(val); }
```

**(a)** In Rust, application developers must choose a concrete container implementation with potentially surprising performance implications.

**(b)** Using Primrose, developers describe the container's expected behaviour via *properties* and the best valid implementation is selected.

■ **Figure 1** Selecting the unique elements of a sequence by inserting the elements into a *set*.

languages under different names, such as, *traits* (e.g., in Rust and Scala), *protocols* (e.g., in Swift), *interfaces* (e.g., in Java), and *type classes* (e.g., in Haskell). All these techniques allow developers to use multiple concrete implementations, such as HashSet and BTreeSet, with a single abstract type, which we might call Set. However, these types are deliberately *abstract*, meaning that we *cannot* instantiate them directly: When creating such a type, a developer must commit to a specific concrete implementation, requiring the developer to look underneath layers of abstraction to make an informed decision. Thus, these abstraction techniques do not free developers from considering low level details and they are not powerful enough to express *semantic* requirements: developer cannot specify their functional requirements directly, but merely provide a common *syntax* enabling the use of multiple implementations. With such an abstract container type Set, we cannot express that each concrete implementation is required to contain no duplicate elements. Similarly, with an abstract type Stack, we cannot state that the last-in-first-out property is respected by the push and pop operations.

Figure 1b shows the same problem of selecting unique elements, but expressed using Primrose. Application developers specify their functional requirements—in this case, that the container must contain unique elements—as a *semantic property*. This semantic property is expressed in lines 1–3 in the Primrose specification language as a logical predicate written as a lambda expression. The property is used to *refine* the container data type in lines 4 and 5. Refinement types have long been used as a technique for program verification—including container types [36]. Here, we use refinement types in a new way, allowing programmers to express the expected behaviour of a container, and freeing them from having to make a (potentially difficult) implementation choice. The remaining code remains unchanged: we can simply use the refined type in line 7. Primrose preprocesses the code from Figure 1b, identifies all valid container implementations from a library of containers, and generates a program equivalent to Figure 1a with the best container implementation inserted automatically.

But which is the *best* container implementation? This depends on the non-functional requirements of the application: Often developers care about fast runtime performance, also, for example, an application might require a low memory footprint. Figure 2 shows the performance and memory consumption for three different implementation choices. Perhaps surprisingly, a custom `UniqueVec` implementation that uses a vector and lazily ensures that the stored elements are unique, by sorting the vector and removing duplicates on access, outperforms the Rust built-in containers `HashSet` and `BTreeSet`. In addition, it is also the best choice for machines with limited memory. Choosing the best container implementation is not always straightforward, particularly as theoretical complexity of operations can sometimes be misleading in the presence of practical effects such as cache-friendliness. `Primrose` selects implementations satisfying developers' functional requirements and opens up opportunities to automatically choose the most desired implementation according to non-functional requirements.

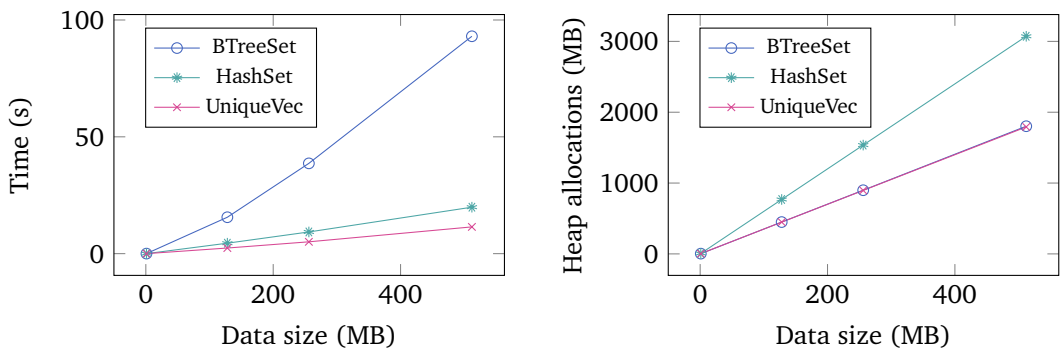

**Figure 2** Runtime performance (left) and memory consumption (right) of three container implementations for storing unique elements of an input sequence from Figure 1a. The custom `UniqueVec` implementation ensures elements to be unique lazily on access. It is the fastest implementation, outperforming `HashSet` and `BTreeSet` from the Rust standard library, while consuming less memory than `HashSet`.

## 3 Overview

Figure 3 gives an overview of the design of the `Primrose` selection tool. Using `Primrose`, the application developer writes code in terms of an abstract type, and a *property specification* describing the syntactic and semantic properties they expect this type to satisfy. The syntactic properties take the form of traits and the semantic properties take the form of type refinements. To write a program, the developer only specifies what functional properties must be satisfied by the required container, and does not have to commit to a particular implementation. In Figure 3, the developer specifies that they require a container (the syntactic property `ContainerT`) where all elements are unique (the semantic property `unique`). We discuss properties in detail in Section 4.

Given this code as input, `Primrose` will, acting as a preprocessor, generate copies of the input code where the abstract type is instantiated into a valid concrete implementation that satisfies the expected properties. It determines which implementations are valid by consulting *library specifications*, which are provided by library developers. These specifications abstract over concrete container implementations and provide a summary of their externally observable semantics. For each implementation, the

**Primrose: Selecting Container Data Types by Their Properties**

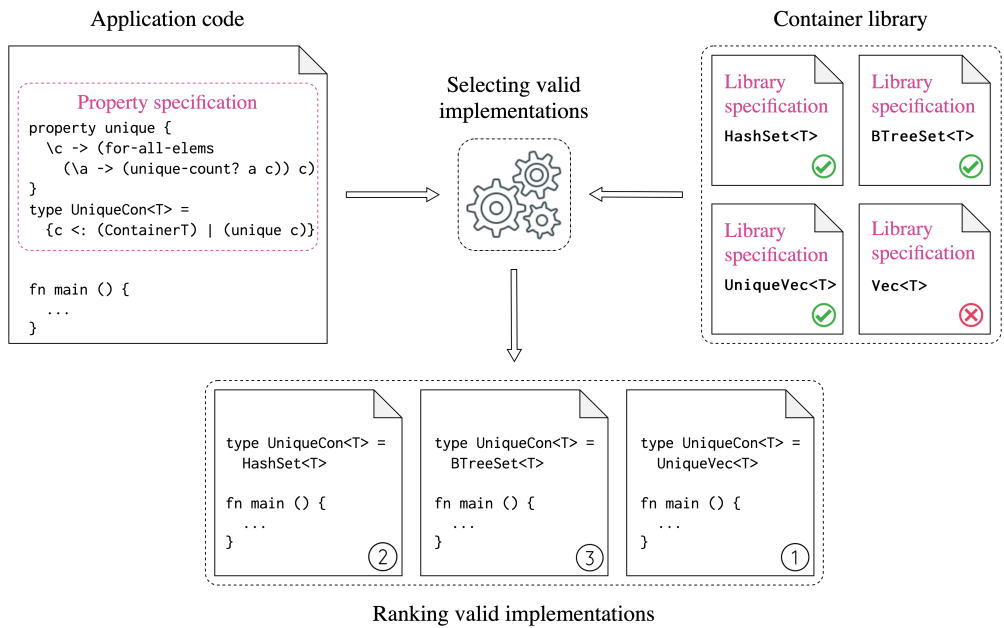

**Figure 3** The workflow of Primrose: *Property specifications* (top left), written and used by the application developer, are used to check which *library specifications* (top right), written by library developers, satisfy them. Valid implementations (marked with a green check marks), are then ranked by their performance (bottom).

library specification contains the pre- and post-conditions of each operation in terms of an abstract list model. We discuss these specifications in more detail in Section 5.

In our example in Figure 3, the library specification of the `Vec<T>` type indicates that it is not a suitable choice for `UniqueCon<T>`, as it does not satisfy the required semantic property `unique`. We use a satisfiability modulo theories (SMT) solver for the selection process, which we discuss in Section 6.

Figure 3 shows at the bottom a simplified version of the generated programs. In our implementation, we ensure that only the container operations that the application developer specifies with syntactic properties are accessible in the generated program. Our current prototype of Primrose focuses on ensuring the functional correctness of selecting container implementations based on desired properties. Nevertheless, we have implemented a simple process that ranks valid implementations by their runtime performance. Rankings by other non-functional metrics could easily be added to our design. We provide discussion about code generation and ranking in Section 6.4.

**Using Rosette as the Common Language for Specifications and Selection** The "solver-aided programming language" Rosette [34, 35] is used as the common language in Primrose for the formal parts. Rosette is chosen for Primrose due to its convenient interface to the Z3 SMT solver and the straightforward translation from Primrose property specifications into Rosette. Property specifications are used as verification conditions when selecting implementations by checking against library specifications which are directly encoded in Rosette. The selection process is done by interacting with the SMT solver via Rosette.

**Portability of Primrose**   Currently, we choose Rust as the target language to implement our idea. Application developers write the property specifications as a part of their Rust programs and Primrose generates Rust code after processing specifications. However, Primrose could easily be ported to many other languages, since property specifications, library specifications, and the process of selecting implementations are all language-agnostic and not attached to Rust's particular type system or language features. Adapting property specifications into other languages only requires such languages to have a construct similar to Rust's traits, such as traits in Scala and interfaces in Java, allowing us to model syntactic properties. It would be straightforward to add new backends to Primrose to generate code in these languages. Our library specifications are, by design, an abstraction over implementation details, describing the intended semantics of container operations without respect to their implementation. This means we can trivially adapt these specifications to container libraries from other languages, so long as our specifications remain an abstraction of the new implementations. Thus, we anticipate that Primrose could easily be adapted to produce code in any language with sufficient support for data abstraction, such as Java, Scala, Swift, or C++.

## 4  Property Specifications

The application developer specifies the desired behaviours of their required container with a *property specification*, for example, for the type UniqueCon from Figure 3:

```
1 property unique { \c -> (for-all-elems (\a -> (unique-count? a c)) c)  }
2 type UniqueCon<T> = {c <: (ContainerT) | (unique c)}
```
Primrose

We first define the *semantic property* unique using a *predicate*. In our specification language, such predicates have type $Con\langle\tau\rangle \to Bool$, where $Con\langle\tau\rangle$ is a placeholder that is resolved into a concrete container type by the selection process. The combinator for-all-elems is part of a library enabling to write predicates for individual container elements. The predicate unique-count? holds iff the given element occurs exactly once in the container. These combinators and predicates are explained in Section 4.2.

With the defined *semantic property* unique, we can then declare the container type UniqueCon<T>. The first part of the declaration specifies the syntactic property that must be satisfied by the container type, in the form of the trait ContainerT. Specifically, c <: (ContainerT) says that the type of the container c must implement the trait ContainerT, which specifies a set of basic container operations. The second part of the declaration *refines* our container type by the predicate unique, stating that the property must be invariant across all container operations. Properties may also be composed. For multiple syntactic properties, we specify a list of traits (c <: (T1, T2)) that the container type implements. For multiple semantic properties, we use conjunction, i.e. ((p1 c) and (p2 c)).

Figure 4 shows the syntax of the Primrose property specification language. Formally, the specification language is a variant of the polymorphic $\lambda$-calculus [15, 27], with restrictions on the use of polymorphism to enable implicit type inference [17, 24]. This type system guarantees termination, making specifications easier to analyse and straightforward to translate into SMT verification conditions in Rosette. The translation into Rosette is straightforward, as terms in the Primrose property specification

**Primrose: Selecting Container Data Types by Their Properties**

language (literals, variables, lambdas, and function application) are translated into their counterparts in the functional Rosette language.

### 4.1 Syntactic Properties as Traits

In our Primrose prototype, we encode syntactic properties as Rust traits, specifying the operations needed by the application developer to interact with a container. Traits are defined in Rust and lifted into our property specification language. For instance, the trait ContainerT introduced above is implemented as:

```Rust
pub trait ContainerT<T> {
  fn len(&self) -> usize;
  fn contains(&self, x: &T) -> bool;
  fn is_empty(&self) -> bool;
  fn insert(&mut self, elt: T);
  fn clear(&mut self);
  fn remove(&mut self, elt: T) -> Option<T>;
}
```

By writing `c <: ContainerT`, the application developer indicates that they expect the container type selected by Primrose to include implementations for all operations in the trait ContainerT. Thus, after executing Primrose, UniqueCon<T> will be resolved into a concrete container type that implements the trait ContainerT.

As mentioned, we can also declare a container type that satisfies multiple syntactic properties. For instance, suppose that in addition to ContainerT, we would like our container to also satisfy the syntactic property IndexableT:

```Rust
pub trait IndexableT<T> {
  fn first(&self)   -> Option<&T>;
  fn last(&self)    -> Option<&T>;
  fn nth(&self, n: usize) -> Option<&T>;
}
```

With just ContainerT, there is no way to observe the *ordering* of elements in the container, but with IndexableT there is, as we can now select elements based on their position. By composing our new syntactic property IndexableT with ContainerT we can now specify a container of unique elements where the order can be observed:

```Primrose
type UniqueIndexableCon<T> = {c <: (ContainerT, IndexableT) | (unique c)}
```

$$
\begin{aligned}
\text{Literals} \quad & l &:=&\ true\ |\ false \\
\text{Terms} \quad & t &:=&\ l\ |\ x\ |\ \lambda x.\ t\ |\ t\ t \\
\text{Refinement} \quad & r &:=&\ t\ |\ r \wedge r \\
\text{Container Type Declarations} \quad & c &:=&\ \{v\ <:\ B\ |\ r\} \\
\text{Simple Types} \quad & \sigma &:=&\ Bool\ |\ T\ |\ Con\langle\sigma\rangle \\
\text{Types} \quad & \tau &:=&\ \sigma\ |\ \tau \to \tau\ |\ \forall T <:\ B.\ \tau \\
\text{Bounds} \quad & B &:=&\ trait\_name\ |\ B\ ,\ B
\end{aligned}
$$

**Figure 4** The syntax of property specifications. $T$ is the type variable, ranging over element types of the target language, which is Rust in this case.

11:8

Semantic properties, such as unique, must be invariant across all operations from all syntactic properties required of the container.

## 4.2 Semantic Properties as Predicates

As mentioned, semantic properties are predicates that are used to construct refinements for container types; each declared container type in the form $\{v <: B \mid r\}$ is a *refinement type*, i.e. a type circumscribed by a logical predicate [13]. When the predicates are in SMT-decidable logic, they can be statically checked [5]. Such techniques are used in programming languages like Liquid Haskell and F*, where they are used to facilitate verification of program correctness. For instance, in Liquid Haskell, we may define a refinement type UniqueList representing a list of unique elements as:

```
{-@ measure unique @-}                                      Liquid Haskell
unique :: (Ord a) => [a] -> Bool
unique [] = True
unique (x:xs) = unique xs && not (S.member x (elts xs))
{-@ type UniqueList a = {v:[a] | unique v} @-}
```

While our syntax for type refinements strongly resembles Liquid Haskell, our refinement types are slightly different, and serve a different purpose. Firstly, Liquid Haskell's refinements are attached to a *concrete type*, in this case a list (written [a]), whereas our refinements are attached to an abstract container type, which is then resolved by Primrose into a concrete implementation. Secondly, Liquid Haskell uses type refinements for the purpose of *correctness*: If a list is declared to have type UniqueList, the Liquid Haskell verifier will check that it satisfies the predicate unique. For example, it will report an error at compile time if given a list that contains duplicates.

```
{-@ notUniqueList :: UniqueList Int @-}                     Liquid Haskell
notUniqueList::[Int]
notUniqueList = [3, 1, 2, 3]
```

Our work instead uses type refinements to specify the semantic requirements of the application developer to guide selection of valid concrete implementations. Once all valid implementations have been found, Primrose simply selects the implementation providing the best performance for the application developer. In short, rather than to aid verification, we use refinement types to help application developers optimise their programs. We give more details on the selection process in Section 6.

**Combinators and Predicate Functions**  Demonstrated by our examples, Primrose provides a set of combinators and predicate functions to facilitate writing of property specifications. These combinators and predicate functions are defined in Rosette and then imported into our property specification language. In the semantic property unique, the combinator for-all-elems is used to specify that the predicate unique-count? must hold for all elements inside the container. The type of the combinator for-all-elems is $Con\langle\tau\rangle \to (\tau \to Bool) \to Bool$, meaning this combinator takes in two arguments, the first of which is a container and the second of which is a predicate on the elements of that container, and eventually returns a boolean value.

For the purposes of checking, we represent containers $Con\langle\tau\rangle$ abstractly in Rosette as lists. We discuss this list abstraction and justify it in Section 5. This means that we can implement our for-all-elems combinator straightforwardly with a list fold operation:

```
1 (define (for-all-elems c fn)                                           Rosette
2   (foldl  elem-and #t (map (lambda (a) (fn a)) c)))
```

We also provide some combinators for applying *relations* between elements in a container. For instance, for-all-consecutive-pairs:

$$\text{for-all-consecutive-pairs} \; : \; Con\langle\tau\rangle \rightarrow (\tau \rightarrow \tau \rightarrow Bool) \rightarrow Bool \tag{1}$$

Unlike for-all-elems, this combinator is given a binary relation between elements, and checks that this relation holds between any two consecutive elements in our container.

With this combinator and the predicates geq? and leq?, we can define properties like ascending and descending, which specify particular orderings of elements in a container:

```
1 property ascending { \c -> (for-all-consecutive-pairs c leq?) }        Primrose
2 property descending { \c -> (for-all-consecutive-pairs c geq?) }
```

Besides the set of combinators and predicate functions predefined in Primrose, application developers may also provide customised functions by providing Rosette definitions and importing them into our property specification language.

**Composition of semantic properties**  As shown in Figure 4, we can compose semantic properties in a container type declaration with conjunction. For instance, to declare a container type with elements arranged in *strictly* ascending order, i.e., both unique and ascending properties must hold, we can write the following:

```
1 type StrictlyAscendingCon<T> = {c <: (ContainerT) | ((unique c)  and (ascending c))}    Primrose
```

This conjunction is straightforwardly translated into a conjunction operation in Rosette.

### 4.3 The Interaction between Semantic and Syntactic Properties

All semantic properties we have seen so far have been invariants across all operations, but some semantic properties relate to specific operations given by syntactic properties. For instance, when specifying a stack container type providing operations push and pop with the expected last-in-first-out (LIFO) property. Firstly, we define a trait specifying operations push and pop, namely StackT:

**Listing 1**  The trait StackT specifying operations push and pop

```
1 pub trait StackT<T> {                                                  Rust
2   fn push(&mut self, elt:  T);
3   fn pop(&mut self) -> Option<T>;
4 }
```

Secondly, we define the semantic property lifo for containers that implement StackT:

**Listing 2**  The semantic property LIFO

```
1 property lifo { \c <: StackT -> (forall  \x. pop (push c x) == x) }    Primrose
```

Unlike previously, this semantic property includes a requirement that the given container implements the trait StackT, enabling us to refer to the operations pop and push inside the semantic property. In this definition, forall is a combinator with type:

$$\text{forall} \;:\; \forall x.\, (x \rightarrow Bool) \rightarrow Bool \tag{2}$$

This combinator is implemented with the forall procedure defined in Rosette's library, which serves as a construct for creating universally quantified formulae.

Armed with the trait StackT and the semantic property lifo, we can combine all these elements and declare our stack type as follows:

```
1  type StackCon<T> = {c <: (ContainerT, StackT) | (lifo   c)}                         Primrose
```

In the next section, we will discuss how library developers write specifications for their container implementations.

## 5 Library Specifications

*Library specifications* abstract over Rust implementations, providing a clear definition of intended semantics of each operation, without respect to performance or implementation details. This approach allows us to select container implementations by simply checking their library specifications, rather than their full implementations, against the properties specified by the application developer. Moreover, using specifications which are abstracted from implementations makes Primrose easy to repurpose for programming languages other than Rust, as the same specifications would apply, with minimal or no modification, to container libraries written in any other language.

By encoding these specifications into *property based tests*, which validate container implementations against their library specifications (Section 7.1), we ensure the selected implementations indeed satisfy a required property specification. Since these library specifications form a *functional correctness* specification for each operation, they could also be used in future as the basis of full functional correctness verification with a verification framework for Rust [23], but this is out of scope for Primrose.

### 5.1 The Basic Design of Library Specifications

Library specifications of concrete container implementations are developed based on Hoare logic [18]. For each concrete container implementation, we provide a set of *Hoare triples*, one for each operation. A Hoare triple of the form $\{\phi\}$ op $\{\psi\}$ states that if the *precondition* $\phi$ holds and the operation op is executed, then the *postcondition* $\psi$ will hold. These conditions are predicates on the state of the program. In our case, the state contains the container, plus any other inputs and outputs of the operation op.

As mentioned in Section 3, we model the container as a list in Rosette for Primrose's library specifications. The list is a model to convey the intended semantics, and does not proscribe anything about the implementation — the implementation is free to represent data in any chosen structure. For example, a set data type may be implemented with a binary search tree, but will still be specified with a list.

**Primrose: Selecting Container Data Types by Their Properties**

These model lists are a simple abstraction, easy to analyse, with which all container operations can be specified.

**Library Specifications Convey the Intended Semantics for Implementations** It is important that all possible executions of a concrete implementation should be captured by its library specification. Otherwise in the process of selecting implementations by checking if their library specifications match the required semantic property, Primrose could select an unsatisfying implementation. More formally, a proof of functional correctness of an implementation w.r.t. its specification would take the form of a data refinement [28], where each value of the concrete container type is related to our list model by an *abstraction function* $\alpha$, and our specification on lists is shown to contain all possible behaviours of the concrete implementation using a *forward simulation*:

$$\alpha^{-1}; \mathsf{op}(C) \subseteq \mathsf{op}(A); \alpha^{-1}$$

(where ; is forward composition of relations and $\alpha^{-1}$ is the inverse relation of $\alpha$)

Here, $\mathsf{op}(C)$ denotes the concrete implementation of our operation $\mathsf{op}$, represented as a relation from inputs to outputs. The abstract operation $\mathsf{op}(A)$ is the maximal relation satisfying the Hoare triple given in our library specification, and $\alpha$ is a suitable abstraction function that flattens a concrete container into a list.

If a forward simulation is shown for all operations, we can conclude that each possible execution with the concrete container has a corresponding execution with an abstract list, thus the specification accurately captures the implementation's semantics.

For instance, a binary search tree $T$ can be abstracted to a sorted list $L$ by an abstraction function *inorder* that does an in-order traversal. For each operation interacting with $T$, there exists a corresponding operation at the abstract level defined using $L$. Take the operation $\mathsf{insert}(T, \mathsf{x})$, which inserts an element $x$ into a binary search tree $T$. We can abstract such an operation to $\mathsf{insert}(L, \mathsf{x})$ which inserts $x$ at the right location in a sorted list. The relation between these two operations is shown by this diagram:

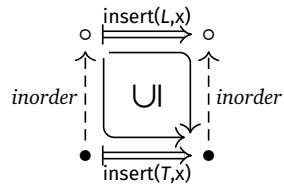

In this work, we specified four container implementations from Rust's standard library (Vec, LinkedList, HashSet, BTreeSet) and four custom container implementations (SortedVec, LazySortedVec, UniqueVec, LazyUniqueVec) by abstracting them into a list model. As we discuss in Section 5.5, library specifications abstract over some implementation details, and, thus, Vec and LinkedList share the same specifications, as do the eager and lazy SortedVec and UniqueVec implementations. For each specification, we define a suitable abstraction function for forward simulation which, while not needed for selection, is used for property-based testing to justify that a concrete implementation satisfies the intended semantics described by its library specification.

**Completeness of Library Specifications** While it is important to ensure that library specifications indeed convey the intended semantics of the implementation, *completeness* of library specifications is also important. Without completeness, Primrose could possibly rule out perfectly valid implementations because it cannot prove that the required semantic properties are preserved for an incompletely-specified operation.

Our approach easily ensures completeness when each operation is specified by a *deterministic* model operation. Forward simulation states that every execution of the concrete implementation has a corresponding execution in the abstract operation, while determinism states that such correspondence is one-to-one, i.e., each abstract execution also has a corresponding concrete one. Thus, just as forward simulation states that each property established for an abstract operation applies also (via the inverse of the abstraction function $\alpha^{-1}$) to a concrete implementation, completeness states that each property established for a concrete implementation applies (via the abstraction function $\alpha$) to the abstract operation. With both completeness and forward simulation, we ensure that *all* valid implementations and *only* the valid implementations are selected by Primrose.

There are many other availiable approaches for modelling library specifications, for instance, the axiomatic approach used in algebraic specifications for abstract data types [39], specifying the behaviour of operations as a set of equational axioms that relate various operations. However, it is hard to ensure the completeness of algebraic specifications, as it is hard capture all behaviours of all operations by a set of equations.

## 5.2 The Library Specification of A LinkedList

Rust's LinkedList is a doubly-linked list. The abstraction function to convert it into a logic list is straightforward: Collect all nodes' values with previous and next pointers.

Firstly, we specify the insertion operation, LinkedList::insert, whose type signature is:

```rust
fn insert(&mut self, elt: T) {…}
```

Since variables in Rosette are immutable, in the corresponding abstract insertion operation, we alter the type to return a new list instead of altering the list in-place[1]:

```
abs-insert: List<T> -> T -> List<T>
```

We can then provide the specification of LinkedList::insert with respect to its corresponding abstract operation, the maximal relation satisfying the Hoare triple:

$$\{xs_0.\ \text{true}\}\ \text{abs-insert}\ \{xs_0\ x\ xs.\ xs = \text{model-insert}\ xs_0\ x\} \quad (3)$$

Here, $xs_0$ refers to the initial value of the container and $xs$ to the resultant container, and $x$ is the element we insert. The function model-insert is defined in Rosette on lists:

```
(define (model-insert xs x) (append xs (list x)))
```

The postcondition states that we expect applying the insertion operation to a container to produce the same result as the model-insert function. In library specifications, defining such *model operations* is a common technique to simplify writing postconditions.

---

[1] Rosette is untyped, but this is morally the type signature.

Similarly, we also provide the specification for the operation `LinkedList::contains`:

```Rust
1  fn contains(&self, x:  &T) -> bool {…}
```

In our corresponding abstract operation, in addition to the boolean value indicating whether the given element `x` is present or not, the input container is also returned, as we would like to express the input container is not mutable, its value remains unchanged after this operation. Also, since the underlying value with type `T` is given by an immutable reference `&T`, in the abstract operation we treat the immutable reference `&T` as simply `T`. The signature of the abstract operation is shown below:

■ **Listing 3** The signature of the abstract operation corresponding to `LinkedList::contains`

```Rosette
1  abs-contains: List<T> -> T -> (List<T>,  bool)
```

The Hoare triple that serves as the specification of `LinkedList::contains` is:

$$\{xs_0. \text{ true}\} \text{ abs-contains } \{xs_0 \; x \; xs \; r. \; (xs, r) = \text{model-contains } xs_0 \; x\} \tag{4}$$

Note that in this specification, the model operation `model-contains` defined in listing 4 has the same type signature as the abstract operation shown in listing 3. It also returns a pair of values: the output list, which is always equal to the input list, and a boolean value indicating if the element is present in the list.

■ **Listing 4** The model operation for checking an element's containment

```Rosette
1  (define (model-contains xs x)
2    (cond [(list?   (member x xs)) (cons xs #t)]
3          [else  (cons xs #f)]))
```

Because `model-contains` returns the unchanged list, it specifies that the `LinkedList::contains` operation should not change the list.

The library specification of the list removal operation is slightly more complicated, we use `T?` to denote that a type may be `null` to express Rust's `Option<T>` type, which is the return type of `LinkedList::remove`. The type signature of `LinkedList::remove` is shown below:

```Rust
1  fn remove(&mut self, x: T) -> Option<T> {…}
```

This operation removes the first occurrence of an element from the given linked list and returns it. If the linked list does not contain the element, `None` is returned and the list remains unchanged. The signature of the corresponding abstract operation is:

```Rosette
1  abs-remove: List<T> -> T -> (List<T>,  T?)
```

The model removal operation has the same signature as the abstract operation. We return `null` in Rosette for the `None` case:

```Rosette
1  (define (model-remove xs x)
2    (cond [(list?   (member x xs)) (cons (remove x xs) x)]
3          [else  (cons xs null)]))
```

Again, we return a pair of the resulting list and the element being removed. Then we provide the library specification of `LinkedList::remove`:

$$\{xs_0. \text{ true}\} \text{ abs-remove } \{xs_0 \; x \; xs \; r. \; (xs, r) = \text{model-remove } xs_0 \; x\} \tag{5}$$

To provide a complete specification of `LinkedList`, the library developer must ensure that each operation of the `LinkedList` is specified by a trait, and for each operation in each trait the `LinkedList` implements, specifications similar to the above are provided.

### 5.3 The Library Specification of A BTreeSet

For the `LinkedList` it is intuitive to use a logic list as a model, as they are both lists. However, even for non-linear structures such as trees, we can still use logic lists as a model. Rust's `BTreeSet` is a set implemented using a b-tree. All elements are unique and arranged in ascending order. Thus, our list model of the b-tree is simply a sorted list in ascending order, where uniqueness of elements is preserved. The abstraction function $\alpha$ that converts the `BTreeSet` to our list model is simply an in-order traversal.

Our first example is again the specification of the insertion operation with signature:

```Rust
pub fn insert(&mut self, value: T) {...}
```

The signature of the abstract insert operation on our model lists is the same as for `LinkedList::insert`. The specification of `abs-insert` for `BTreeSet`, however, differs from that of `LinkedList`, as we must maintain ordering and uniqueness of elements:

$$\{xs_0.\ xs_0 = \text{dedup (sort } xs_0 \text{ <)}\}\ \text{abs-insert}\ \{xs_0\ x\ xs.\ xs = \text{model-insert}\ xs_0\ x\} \quad (6)$$

As before, $x$ is the element to be inserted, and $xs_0$ and $xs$ are lists modelling the container (via the in-order traversal function $\alpha$) before and after the `abs-insert` operation respectively. We place an assertion $xs_0 = \text{dedup (sort } xs_0 \text{ <)}$ in the precondition requiring that the model $xs_0$ to be a sorted list of unique elements. While this precondition should always be satisfied by an in-order traversal of a valid b-tree, we do not want our abstraction to constrain the implementation's behaviour if the data invariants of the b-tree are violated — given a malformed b-tree, the implementation should be free to return any result. Because the semantics of `abs-insert` are the maximal relation satisfying this specification, this abstract operation contains all possible behaviours of the concrete implementation if this precondition is violated. The `model-insert` here is simply an insertion operation defined on a sorted list of unique elements:

```Rosette
(define (model-insert xs x) (dedup (sort (append xs (list  x))  <)))
```

We can also provide specifications for abstract operations that observe the ordering of elements in a `BTreeSet`, such as those operations from the `IndexableT` trait, since there is a one-to-one correspondence between each element's position in a `BTreeSet` and its position in the model list abstracted from the `BTreeSet`. For instance, we provide the specification of the operation `BTreeSet::first`, which is the operation obtaining the first (and also the minimal) element of a `BTreeSet` with signature:

```Rust
fn first(&self)   -> Option<&T> {...}
```

We again provide the signature of its corresponding abstract operation:

```
abs-first:  List<T>  ->  (List<T>,   T?)
```

Like `LinkedList::contains` in listing 3, this type includes a returned list, as Primrose does not consider the immutability of `&self` in the Rust type signature above. We again include the requirement that the container is unchanged in the specification:

$$\{xs_0.\ xs_0 = \text{dedup (sort } xs_0 \text{ <)}\}\ \text{abs-first}\ \{xs_0\ xs\ x.\ (xs, x) = \text{model-first}\ xs_0\} \quad (7)$$

Here, `model-first` is defined as a function that returns the first element of the list, is present, along with the list itself:

```
1 (define (model-first xs) (cond [(null? xs) (cons xs null)] [else (cons xs (first   xs))]))                           Rosette
```

As before, our precondition includes the assumption that the model $xs_0$ abstracted from the BTreeSet contains unique elements that are sorted in ascending order.

### 5.4 The Library Specification of A HashSet

A tree implementation of a set maintains its elements in a fixed ascending order, and the ordering of our abstract list model simply reflects the ordering of the elements in the tree. However, some container implementations do not have a fixed ordering of elements. For instance, the HashSet in Rust is a set implementation using a hash algorithm which is randomly seeded. Despite the implementation storing elements in an unspecified order, we may still safely use a sorted, ascending list of unique elements as our abstract model of a HashSet: Our abstraction function $\alpha$ merely collects all elements from the HashSet into a list and then sorts them into ascending order.

Since the ordering of elements in our list is now different from the ordering of elements in the HashSet, the developer may specify properties relating to the ordering of elements, such as ascending, that are not satisfied by the implementation, but are trivially satisfied by the abstraction function. This would lead to HashSet being considered a valid choice for an ascending container. However, Primrose prevents this by the checking of syntactic properties. The HashSet type does not implement any trait with operations that allow the ordering of its elements to be observed.

Therefore, in applications for which the ordering of elements is important, HashSet is never a valid choice. The selection process of valid implementations according to traits is discussed in Section 6.1.

If a library developer decides to write a HashSet with operations that leak ordering, they can provide a nondeterministic library specification for such a HashSet that can still be used by Primrose in the selection process.

For the operations defined on HashSet and BTreeSet, such as insert, remove and contains, the specifications of both implementations are identical—after all, the only observable difference between the implementations is performance—but the specification for HashSet lacks operations that observe the ordering of its elements, such as first or last.

### 5.5 Abstracting Over Implementation Details with Library Specifications

Since the basic container operations of both HashSet and BTreeSet have the same externally observable behaviour, we can use the same specifications for both implementations. There are many such cases where specifications can be re-used: For instance, we provide two implementations of an ascending vector: SortedVec and LazySortedVec. SortedVec maintains the ascending order of elements inside the vector on insertion (*eager*) and LazySortedVec instead sorts elements whenever the vector is accessed (*lazy*). Since both implementations share the same externally observable behaviour, we use the same model for both implementations: A list with elements sorted in ascending order. Also, their operations are specified with the same set of model operations. For the eager implementation, the abstraction function $\alpha$ simply collects all its elements into a list. For the lazy implementation, in addition to collecting all elements into a list, the abstraction function $\alpha$ also sorts elements into ascending order.

## 6 Selecting and Ranking Implementations

Before ranking container implementations by performance or other non-functional metrics, Primrose must first identify all implementations that comply with the property specifications provided by the application developer.

### 6.1 Selecting Container Implementations Satisfying Syntactic Properties

The first step of selecting valid implementations is to select concrete container implementations from the library that satisfy required syntactic properties in a property specification, which is straightforward. Primrose simply picks concrete container implementations that implement the traits required by the property specifications.

For instance, suppose that in a property specification, an application developer requires a container type implementing traits ContainerT and IndexableT, the elements of which are sorted in ascending order:

**Listing 5** Property specification composing properties: ascending, ContainerT and IndexableT

```
1 property ascending { \c -> (for-all-consecutive-pairs c leq?) }
2 type AscendingIndexableCon<T> = {c <: (ContainerT, IndexableT) | (ascending c)}
```
Primrose

Rust's collections library has concrete container implementations Vec, LinkedList, BTreeSet and HashSet, where Vec, LinkedList and BTreeSet implement both required traits while HashSet does not implement the trait IndexableT. Clearly, HashSet does not satisfy all required syntactic properties. Therefore, HashSet is ruled out as a possible implementation for AscendingIndexableCon<T>. The implementation for AscendingIndexableCon<T> is then selected from the remaining Vec, LinkedList and BTreeSet types by checking if the library specifications satisfy the required semantic property, ascending.

### 6.2 Selecting Container Implementations Satisfying Semantic Properties

After gathering container implementations with required syntactic properties, Primrose selects the ones that satisfy the required semantic properties from these candidates. As discussed in Section 5, our library specifications abstract over the concrete container implementations, describing their externally observable semantics in a compact and tractable format. Primrose performs this selection process by encoding the property specifications as verification conditions against the candidates' library specifications in Rosette, to be discharged by an SMT solver in Rosette's backend.

To generate the required verification conditions, Primrose first translates the required semantic properties, given in the specification language of Primrose, into definitions in Rosette that can be used by the solver. The container type Con<T> is resolved into the model type used in our library specifications, i.e., a logic list. For instance, the generated code according to the property ascending from listing 5 is:

**Primrose: Selecting Container Data Types by Their Properties**

```
1  (define ascending (lambda (c) (for-all-consecutive-pairs c leq?)))            Rosette
```

With these Rosette definitions, Primrose generates verification conditions. For example, to check if BTreeSet is ascending, Primrose checks that the semantic property ascending is an invariant held across each operation defined for BTreeSet. For instance, for the insertion operation, specified by (6) in Section 5.3, it checks that the property ascending is preserved by any execution that satisfies its precondition and its postcondition:

$$\forall\, xs_0\; xs\; x.\; \frac{xs_0 = \mathsf{dedup}\;(\mathsf{sort}\;xs_0\; \mathsf{<}) \qquad xs = \mathsf{model\text{-}insert}\; xs_0\; x}{\mathsf{ascending}\; xs_0 \Rightarrow \mathsf{ascending}\; xs}$$

(where: $\exists\, xs_0.\; \mathsf{ascending}\; xs_0 \wedge xs_0 = \mathsf{dedup}\;(\mathsf{sort}\;xs_0\; \mathsf{<})$)

■ **Figure 5** The rule for checking the operation BTreeSet::insert against ascending

Recall that $xs_0$ and $xs$ are model lists abstracted from the BTreeSet, specifically, $xs_0$ is the model for the input BTreeSet, and $xs$ is the model for the resulting BTreeSet of BTreeSet::insert. The model operation model-insert specifies the behaviour of BTreeSet::insert's corresponding abstract operation. Given the rule shown in Figure 5, the solver attempts to find a counterexample, i.e., for all input models $xs_0$ that satisfy the semantic property ascending, the solver tries to find a resulting model of the operation that does not satisfy the property. If there is no such counterexample found, the solver will conclude that the operation BTreeSet::insert satisfies the property ascending.

This search for a counterexample is parameterised by a *model size*, which denotes the maximum size of the input list $xs_0$ considered by the solver. This parameter is configurable by the application developer using Primrose, and its impact on Primrose's selection time is evaluated in Section 7.2.

The rule contains a side condition stating that there should be no contradiction between the required semantic property and the precondition of the operation. This side condition is important for ensuring that the solver does not search for a counterexample in an empty search space then falsely conclude that the absence of the counterexample means that the property holds. The side condition requires that there exists at least one model that satisfies both the precondition of the operation and the required semantic property. Without the side condition, the rule is unsound.

In general, the library specification of each operation takes the form:

$$\{\phi(xs_0, \vec{u})\}\; \mathsf{op}\; \{\psi(xs_0, xs, \vec{v})\} \tag{8}$$

where $xs_0$ is the (abstract list model of the) input container and $xs$ is the result of the operation op. The sets of variables $\vec{u}$ and $\vec{v}$ denote any additional variables involved in the specification, such as additional inputs or outputs to the operation. The general form of the verification condition Primrose generates for the SMT solver, to check if an operation op satisfies a property $P$, is given in Figure 6.

For our BTreeSet example, Primrose checks these verification conditions for each operation of ContainerT and IndexableT—the traits implemented by BTreeSet. Because the property ascending is satisfied by all operations, Primrose concludes that the BTreeSet is a valid implementation choice for the required container type AscendingIndexableCon<T>.

$$\forall\ xs_0\ xs\ \vec{u}\ \vec{v}.\ \frac{\phi(xs_0, \vec{u}) \qquad \psi(xs, \vec{v})}{P(xs_0) \Rightarrow P(xs)} \quad (\text{where: } \exists\ xs_0\ \vec{u}.\ P(xs_0) \wedge \phi(xs_0, \vec{u}))$$

**Figure 6** The rule for checking an operation against a property

The same checks are also run for the other two candidates that satisfy the required syntactic properties (Vec and LinkedList) but they do not satisfy the required semantic property ascending. Therefore, Primrose concludes that only BTreeSet is a valid implementation for the required container type AscendingIndexableCon<T>.

### 6.3 Handling Interactions Between Semantic and Syntactic Properties

In this section, we discuss how Primrose selects library implementations with semantic and syntactic properties, such as the stack container StackCon<T> from Section 4.3, where the operations push and pop specified in the trait StackT (listing 1) are made available to the semantic property lifo (listing 2).

Firstly, Primrose generates the definition of semantic property lifo in Rosette, where the operations push and pop are now replaced with their model operations:

```
(define lifo (lambda (c) (forall (list x) (equal? (cdr (model-pop (model-push c x))) x))))
```
Rosette

The specific model operations model-pop and model-push are supplied to this definition for each candidate type considered by Primrose. Recall that our library specifications state that these model operations exactly specify the intended behaviour of every library operation, which means that these model operations can be used here to express assertions about the interaction between operations such as push and pop. Such assertions will, by virtue of forward simulation, also apply to the concrete implementations of the data type.

To illustrate the selection process, suppose a library developer provides two implementations that implement push and pop. The first one is a last-in-first-out implementation, where the library specification of push and pop is:

$$\{xs_0.\ \text{true}\}\ \text{abs-push}_1\ \{xs_0\ x\ xs.\ xs = \text{model-push}\ xs_0\ x\} \tag{9}$$

$$\{xs_0.\ \text{true}\}\ \text{abs-pop}_1\ \{xs_0\ xs\ x.\ (xs, x) = \text{model-pop}\ xs_0\} \tag{10}$$

And the model operations are defined as:

```
(define (model-push-front xs x) (append xs (list x)))
(define (model-pop xs) (cond [(null? xs) (cons xs null)]
                             [else (cons (take xs (- (length xs) 1)) (last xs))]))
```
Rosette

With these two model operations, the solver can verify that this library specification satisfies the semantic property lifo.

By contrast, the second implementation is a first-in-first-out implementation. The library specification of push and pop appears similar:

$$\{xs_0.\ \text{true}\}\ \text{abs-push}_2\ \{xs_0\ x\ xs.\ xs = \text{model-push}\ xs_0\ x\} \tag{11}$$

$$\{xs_0.\ \text{true}\}\ \text{abs-pop}_2\ \{xs_0\ xs\ x.\ (xs, x) = \text{model-pop}\ xs_0\} \tag{12}$$

However, the model operations have different semantics:

```
1 (define (model-push-end xs x) (append (list x) xs))                                Rosette
2 (define (model-pop xs) (cond [(null? xs) (cons xs null)]
3                              [else (cons (take xs (- (length xs) 1)) (last xs))]))
```

With these two model operations, the solver correctly concludes that this library specification does not satisfy the semantic property lifo, and Primrose does not consider this implementation as a valid choice for the container StackCon<T>.

### 6.4 Code Generation and Ranking Implementations by Performance

Once Primrose has selected the valid container implementations, it will generate a Rust program for each valid candidate by resolving the property specification into the selected container implementation. In Figure 3 we show a simplified version of generated programs where property specifications are directly replaced with concrete implementations, in practice Primrose carefully generates Rust's trait objects to encapsulate the concrete implementation and exposing only those operations in Rust traits which are specified as syntactic properties.

As a proof-of-concept implementation, the current Primrose prototype ranks the generated Rust code for each valid implementation by executing all candidates and measuring their runtime on some test input data. We anticipate adopting more sophisticated ranking techniques, such as the ones discussed in the related work, in the future. Our existing prototype of Primrose focuses on enabling application developers to specify their functional requirements, and automating the selection of valid container implementations.

## 7 Evaluation

For Primrose to be feasible for use as a programming tool, it must be practical to ensure that our library specifications are sound abstractions of our container implementations, and the selection process itself must not take a prohibitively long time. Our evaluation demonstrates feasibility in both of these aspects. All measurements are conducted on a MacBook Pro with 32 GB of RAM and a 2.4 GHz 8-Core Intel Core i9 processor.

### 7.1 Correctness of Library Implementations w.r.t their Library Specifications

To ensure the selected implementations are correct, we validate our Rust container library implementations against the library specifications using property-based testing [10]. We use the framework proptest [2] for encoding and performing the tests.

Firstly, we encode the model list with its operations in Rust. Specifically, we encode the model list from Rosette as an immutable ConsList [32] in Rust, along with all its operations. Then we implement the abstraction function $\alpha$ for each container implementation, and, like Chen et al. [8, 9], we encode the forward simulation obligation for the library specification of each operation as assertions in a test.

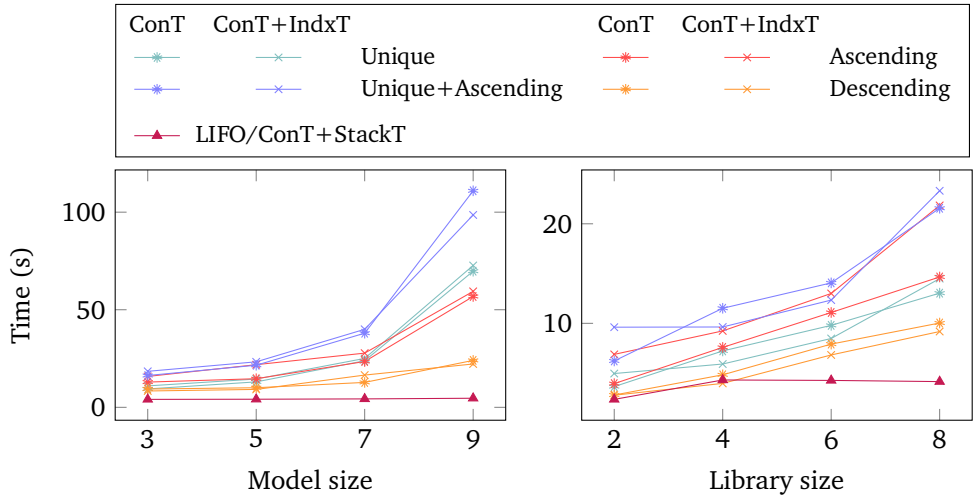

**Figure 7** Primrose's efficiency of selecting implementations for different properties

For each test, 100 test inputs are randomly generated. For our library with eight container implementations, in total 7200 inputs are tested in 7.315 seconds. We conclude that with the existing testing framework, we are able to validate the functional correctness of our container implementations w.r.t. our library specifications efficiently, ensuring that implementations selected by Primrose are correct.

## 7.2 Evaluation of Primrose's Selection Time

For Primrose to be practical, it must perform selection with a reasonable time, even though, as a pre-processing tool, it does not have to be invoked on every compilation run. After the initial invocation, it will only be invoked if the property specification or any library specifications are changed.

The efficiency of the SMT-based selection time is mainly determined by two factors: the *model size* and the *library size*, which together define the search space in which the solver attempts to find a counterexample. If a counterexample is found, Primrose will conclude that the library specification does not satisfy the required semantic property. We expect the solver time to grow linearly with the number of container implementations from which we select (library size) and non-linearly with the model size, which is the length of the input model list to the abstract operation, as this should grow the search space polynomially.

Figure 7 shows the measurements of Primrose's selection time. The left side shows that the selection time, for a fixed library size of eight implementations, increases with the model size. The right side shows that for a fixed model size of five, the selection time increases linearly when the library size is increased.

The complexity of the property specifications and the number of satisfying implementations are also factors that affect the efficiency of the selection, since they determine how difficult it is for the solver to find a counterexample. For example, since the definition of lifo has constant complexity, the model size and library size do not affect its selection time as much as for properties with high polynomial complex-ity such as `unique` and `ascending`. None of our example containers satisfy the property `descending`. As SMT solvers are faster at finding a counterexample than exhaustively proving that no counterexample exists, the selection time for `descending` is faster than for `ascending`, despite both properties having the same algorithmic complexity.

The selection is always completed in less than 30 seconds with a model size of 3 and the full library of 8 implementations. Thus, we consider Primrose to be a practically feasible tool. An increase in model size raises selection time quickly, but in practice, a model with size of more than five is not required to admit counterexamples for most conceivable semantic properties that the application developer may write. This is based on the small scope hypothesis in Alloy [20]. The experimental results show, that Primrose can easily be used for medium-size libraries.

## 8 Discussion of Limitations

Primrose's prototype implementation has some limitations that we discuss here.

Primrose currently covers properties of sequential containers like lists and sets, and we have not yet looked into associative containers like maps and dictionaries. However, we believe it should be possible to characterise them with the same technique: application developers using syntactic properties to describe desired operations and semantic properties to state predicates that should be held by keys and values, and library developers providing library specifications using a list model with key-value pairs as elements.

The other limitation of our current implementation is that we implicitly require all elements inside a container to have some ordering for them to be comparable using `leq` and `geq`. In the future, we should allow application developers to state if the elements inside the container are comparable or have ordering by enriching the syntax and type system of our property specification language.

## 9 Related Work

**Refinement types** Refinement types, first introduced for ML [13], are types enriched with logical predicates, often from an SMT-decidable logic [5], allowing programmers to express rich logical constraints in the type system and automatically check them. Refinement types have recently been implemented in languages such as Haskell [36, 37] and F* [33], supporting very rich specifications suitable for verifying the correctness of programs. While the syntax of Liquid Haskell inspires our design of the syntax of property specifications, we use refinement types not for verification, but for data abstraction, allowing application developers to specify their semantic requirements for the selection process.

**Abstract data types and formal methods** Existing work in algebraic specifications [16, 39] provide a formal definition of abstract data types where the semantics of operations are specified with a set of equational axioms. By contrast, our library specifications are model-based. As mentioned in Section 5.1, this allows us to easily ensure completeness of library specifications. There exist many formal modelling tools that facilitate model-based specification of abstract data types and software systems more generally, for example Z [31], VDM [21], and most recently Alloy [19]. While these tools allow application developers to formally analyse and explore the software design space, including formal reasoning about abstract data types, they work purely on the level of models and do not typically connect to actual code, as Primrose does.

**Performance-oriented selection techniques** Many techniques for design space exploration, particularly machine learning techniques, have been applied in compilers [14] to selected performance optimization techniques [6] using various characteristics as features that are then used to rank the performance of multiple implementations [30]. Many dynamic selection techniques have been developed for assisting the selection of performant containers, based on different evaluation criteria such as workload data [12], architectural concerns [22] and runtime metrics [29]. In addition to dynamic container selection, CoCo [40] is tool allowing safe online switching. None of these techniques, however, provide a general scheme to allow application developers to specify desired behaviour, instead, they purely focus on selecting between multiple, pre-known, valid container implementations. Such techniques could be incorporated into Primrose's ranking process, and are highly complementary with our work.

## 10  Conclusion

We have applied techniques from verification and formal methods in a new way, raising the level of abstraction by freeing developers from the burden of choosing concrete container implementations. Instead, application developers can specify their expected behaviour using semantic properties—a highly general abstraction technique. We provide a methodology to specify container libraries with library specifications, and describe our mechanism to check semantic properties against these specifications using SMT solvers. We implement Primrose for Rust and specify eight Rust container implementations. We show that Primrose is a practical tool that can be feasibly integrated into a programmer's workflow.

**Data Availability Statement**    The artifact of this paper is on Zenodo [26].

**Acknowledgements**    The author Xueying Qin would like to express her special thanks to FromSoftware. Their game Elden Ring provided her remarkable mental supports through the development of the prototype of Primrose. Also, she would like to express her gratitude to HAL Laboratory, their game Kirby and the Forgotten Land provided her aesthetic inspirations for designing the interface of Primrose and served as her great company when she was writing this paper. Lastly, she would like to thank ATLUS. Their game Persona 5 Royal gives her courage to deal with obstacles in the past two rounds of submissions and revisions.

## About the authors

**Xueying Qin** is a PhD student at the University of Edinburgh. Contact her at xueying.qin@ed.ac.uk.

**Liam O'Connor** is a lecturer in Programming Languages for Trustworthy Systems at the University of Edinburgh. His research focuses on combining formal methods techniques with practical programming languages and tools. Contact him at l.oconnor@ed.ac.uk.

**Michel Steuwer** is a lecturer in Compilers and Programming Languages at the University of Edinburgh. His research on compiler design and domain-specific languages aims to drastically simplify the programming of complex parallel hardware devices while improving performance and efficiency. Contact him at michel.steuwer@ed.ac.uk.